\documentclass[10pt, 3p]{elsarticle}

\usepackage[english]{babel}
\usepackage[utf8]{inputenc}
\usepackage{graphics}
\usepackage{graphicx}
\usepackage{amsmath,amssymb,esint}
\usepackage[colorlinks]{hyperref}
\usepackage[x11names]{xcolor}

\biboptions{sort&compress,authoryear}

\newcommand{\reviewertwo}[1]{{\color{black}{#1}}}

\begin{document}

\begin{frontmatter}

\title{Cavitation onset in transient pressure fields}

\author{Pierre Coulombel}
\author{Fabian Denner\corref{cor1}}
\ead{fabian.denner@polymtl.ca}
\address{Department of Mechanical Engineering, Polytechnique Montréal,\\ Montréal, H3T 1J4, Québec, Canada}
\cortext[cor1]{Corresponding author: }

\begin{abstract}
While it is well known that cavitation occurs in liquids under tension, no universally accepted criterion for its onset in transient pressure fields exists. We propose a precise definition of the critical tension for cavitation in transient pressure fields that bridges the gap between quasi-static and dynamic regimes, identifying cavitation as the transition of the bubble radius to a dynamically unstable state. This threshold depends on the instantaneous state of the gas-liquid system and, when combined with an appropriate set of dimensionless parameters, yields a self-similar description of cavitation onset. Phase maps for different liquids reveal a minimum tension required for the onset of cavitation, determined by the duration of the tension event and the initial bubble size, whereby the well-known Blake threshold is the lower bound for cavitation across all conditions.
\end{abstract}

\begin{keyword}
Cavitation \sep Bubble dynamics\\~\\
\textcopyright~2025. This manuscript version is made available under the CC-BY 4.0 license. \href{http://creativecommons.org/licenses/by/4.0/}{http://creativecommons.org/licenses/by/4.0/}
\end{keyword}

\end{frontmatter}

\section{Introduction}
\label{sec:introduction}

Cavitation generally refers to the radial dynamics of gas and vapour bubbles driven by pressure variations, for instance as a result of changes in flow speed (cf.~hydrodynamic cavitation) or due to pressure waves (cf.~acoustic cavitation). Although it is understood that cavitation occurs if liquids are put under sufficient tension or, more generally, the pressure falls below the liquid vapour pressure for a sufficient amount of time \citep{Brennen1995, Caupin2006}, a precise definition of cavitation and the conditions for its onset are currently not available in the literature.

A widely used criterion to define the onset of cavitation is the Blake threshold \citep{Neppiras1951}, which corresponds to the critical tension at which a gas bubble becomes unstable, as surface tension can no longer counterbalance the internal gas pressure and ambient tension, leading to unbounded growth. 
However, the Blake threshold is obtained under the assumption that the liquid pressure changes slow enough, such that the bubble remains in mechanical equilibrium at all times.
This quasi-static assumption is insufficient to describe cavitation onset in transient pressure fields \citep{Prosperetti1984a} or when bubbles interact with each other \citep{Ida2009, Coulombel2024}.
Dynamic thresholds for the onset of cavitation \citep{Flynn1978a, Neppiras1980, Apfel1981} or ad-hoc estimates \citep{Apfel1991} have since been presented, but they only allow to predict a possible degree of cavitation activity.
Several \textit{cavitation numbers} have been proposed to predict cavitation onset in high-speed flows \citep{Plesset1949} or rapidly accelerated liquids \citep{Pan2017, Sobral2024}, yet they do not account for the transient nature of cavitation.
The literature on medical ultrasound typically distinguishes between {\it steady} cavitation (stable oscillations around an equilibrium radius) and {\it transient} or {\it inertial} cavitation (resulting in rapid and often violent bubble collapse), whereby the transition to {transient} cavitation is characterized either by a critical radius $2 R_0 \lesssim R_\text{crit} \lesssim 3.5R_{0}$, where $R_0$ is the initial bubble radius \citep{Flynn1975a, Chomas2001}, or by the Blake threshold as an estimate for the lower bound of the critical tension \citep{Ilovitsh2018, Guemmer2021}. Although many of these definitions for cavitation onset have been successful in predicting cavitation, they are tailored to specific applications or rest on limiting assumptions. 

We propose a precise definition of the critical tension for cavitation onset under transient conditions, marked by the transition of the bubble radius to a dynamically unstable state. Building on this threshold and using a single tension pulse as a canonical transient pressure field, we study the cavitation onset of air bubbles in commonly used liquids and we construct a dimensionless framework that enables a self-similar characterization of cavitation onset.

\section{Methodology}
\label{sec:methodology}

{We consider a spherical bubble that is subject to a single tension pulse. The radial dynamics of this single gas bubble in a Newtonian liquid, assuming spherical symmetry,} are modelled using the Keller-Miksis model \citep{Keller1980},
\begin{multline}
    \left( 1 - \frac{\dot{R}(t)}{c} \right) R(t) \Ddot{R}(t) + \frac{3}{2}\left( 1 - \frac{\dot{R}(t)}{3c} \right)\dot{R}(t)^{2} =  \left( 1 + \frac{\dot{R}(t)}{c} \right)\frac{p_{\text{L}}(R(t)) - p_{\infty}(t)}{\rho} +\frac{{R}(t)}{c} \frac{\dot{p}_{\text{L}}(R(t)) - \dot{p}_{\infty}(t)}{\rho},
    \label{eq:KM}
\end{multline}
where $t$ denotes time, $R$ is the radius of the bubble, $\dot{R}$ is the bubble wall velocity, $\Ddot{R}$ is the bubble wall acceleration, $p_{\infty}$ is the pressure in the far field, $c$ is the speed of sound of the liquid, and $\rho$ stands for the density of the liquid. {Eq.~\eqref{eq:KM} is solved using the  embedded Runge-Kutta RK5(4) scheme of \citet{Dormand1980}.} 

The liquid pressure at the gas-liquid interface is given as
\begin{equation}
    p_{\text{L}}(R(t)) = \left( p_0 + \frac{2\sigma}{R_0} \right) \left( \frac{R_0}{R(t)} \right)^{3\kappa} - \frac{2\sigma}{R(t)} - 4\mu \frac{\dot{R}(t)}{R(t)},
    \label{eq:pL}
\end{equation}
where $\sigma$ denotes the surface tension coefficient, $\mu$ is the dynamic viscosity of the liquid, $\kappa$ is the polytropic exponent of the gas, $p_{0}$ is the reference ambient pressure, and $R_{0}$ is the initial bubble radius. {Three different liquids are considered in this study: water ($\rho = 1000$ kg/$\text{m}^{3}$, $\mu = 1.002 \times 10^{-3}$ Pa.s, $\sigma = 7.28 \times 10^{-2}$ N/m, $c = 1481$ m/s), liquid aluminium ($\rho = 2375$ kg/$\text{m}^{3}$, $\mu = 1.300 \times 10^{-3}$ Pa.s, $\sigma = 8.60 \times 10^{-1}$ N/m, $c = 4600$ m/s), and ethanol ($\rho = 789$ kg/$\text{m}^{3}$, $\mu = 1.074 \times 10^{-3}$ Pa.s, $\sigma = 2.23 \times 10^{-2}$ N/m, $c = 1144$ m/s).}
Without loss of generality, heat and mass transfer are neglected and the vapour pressure of the liquid is $p_\text{v} = 0$.

The bubble is subjected to a single tension pulse, representing a canonical transient pressure field, defined as
\begin{equation}
    p_{\infty}(t) = \begin{cases}
        p_0 - (p_0 - p_{\text{ng}}) \sin^{2}{\left( \pi t/\tau \right)} & t < \tau \\
        p_0 & t \ge \tau,
    \end{cases}
    \label{eq:pulse}
\end{equation}
where $\tau$ is the duration and $p_{\text{ng}} < 0$ is the amplitude of the pulse.
{The response of a single spherical bubble to such a change in pressure is characterized by its inertial timescale \citep{Franc2005},}
\begin{equation}
    t_\text{i} = R_{0} \sqrt{\frac{\rho}{\Delta p}},
    \label{eq:t_i}
\end{equation}
{which follows from the well-known Rayleigh collapse time \citep{Rayleigh1917}. To characterize the onset of cavitation, the relevant pressure difference is defined as $\Delta p = p_{\text{G},0} - p_{\text{ng}}$, where $p_{\text{G},0} = p_{0} + 2 \sigma/R_0$ is the initial gas pressure inside the bubble.}

{All results presented in this paper have been produced with the open-source software library {\tt APECSS} \citep{Denner2023a} and the Zenodo repository accompanying this study (see the data availability statement) contains the code to reproduce the presented results. The simplifications and resulting limitations of the applied methodology are discussed in detail in Section \ref{sec:simplifications}.}

\section{Cavitation threshold}
\label{sec:threshold}

\begin{figure}
    \centering
    \includegraphics[width=0.6\linewidth]{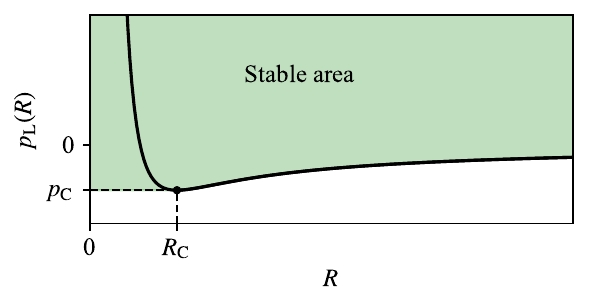}
    \caption{{Schematic illustration of the liquid pressure $p_\text{L}$, as given by Eq.~\eqref{eq:pL}, as a function of the bubble radius $R$, assuming that the evolution of the bubble radius is a quasi-static process, i.e.~$\dot R \rightarrow 0$. The minimum of $p_\text{L}$ is the Blake threshold $p_\text{C}$, given by Eq.~\eqref{eq:p_C}, at the corresponding radius $R_\text{C}$, given by Eq.~\eqref{eq:R_C}.}}
    \label{fig:p_L}
\end{figure}

To establish a general criterion for the onset of cavitation under transient conditions, we hypothesize that cavitation is defined by the bubble radius exceeding a critical value beyond which the bubble is dynamically unstable. As illustrated in Figure \ref{fig:p_L}, Eq.~\eqref{eq:pL} delineates a stable area in the $R$-$p_\text{L}$ plane in which surface tension can counterbalance tension in the surrounding liquid. {Hence, we define cavitation as the bubble radius growing beyond this stable area in which surface tension stabilizes the bubble.}

\subsection{{The Blake threshold}}
\label{sec:sub_Blake}

The minimum of the curve bounding the stable area {in Figure \ref{fig:p_L}} is known as the Blake threshold, {$p_{\text{C}}$, which is a \textit{quasi-static} tension threshold for the onset of cavitation \citep{Neppiras1951}}. If a bubble is put under tension in a quasi-static manner, the bubble radius expands as the liquid pressure decreases, with the Blake threshold being the smallest liquid pressure for which the bubble remains stable. To compute $p_\text{C}$, we first define the Blake radius
 \begin{equation}
    R_{\text{C}} = \left[ \frac{3 \kappa}{2 \sigma} \left( p_0 + \frac{2\sigma}{R_0} \right) R_{0}^{3\kappa} \right]^{\frac{1}{3\kappa - 1}},
    \label{eq:R_C}
\end{equation}
which is the radius corresponding to the minimum of Eq.~\eqref{eq:pL} by assuming that the evolution of the bubble radius is a quasi-static process, i.e.~$\dot R \rightarrow 0$. Evaluating Eq.~\eqref{eq:pL} at the Blake radius $R(t)=R_\text{C}$ with $\dot R = 0$, the Blake threshold follows as
\begin{equation}
    p_\text{C} = \left( p_0 + \frac{2\sigma}{R_0} \right) \left( \frac{R_0}{R_\text{C}} \right)^{3\kappa} - \frac{2\sigma}{R_\text{C}}.
    \label{eq:p_C}
\end{equation}
{Importantly, the Blake threshold assumes that the bubble is in mechanical equilibrium with its surroundings at all time and is based only on the properties of the gas-liquid system, without accounting for dynamic and time-dependent effects. 
Consequently, the Blake threshold cannot represent the onset of cavitation in response to a transient tension event, because Eq.~\eqref{eq:p_C} neither accounts for a finite expansion rate of the bubble ($\dot{R} > 0$) nor for a time-dependent ambient pressure.}

\subsection{{Cavitation threshold in transient pressure fields}}
\label{sec:newthreshold}

\begin{figure}
    \centering
    \includegraphics[width=\linewidth]{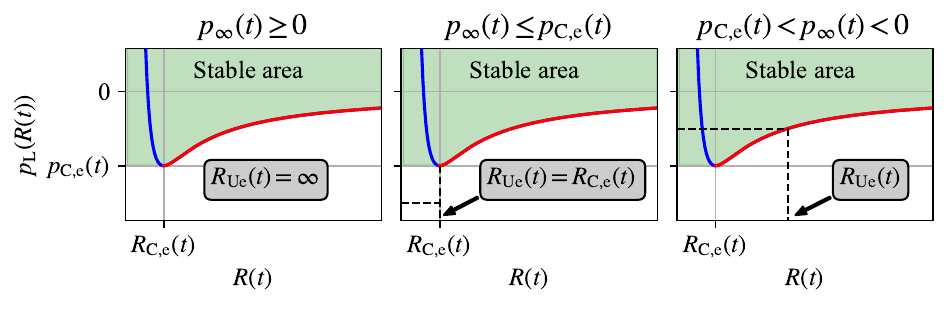}
    \caption{Schematic illustration of the liquid pressure $p_\text{L}$ as a function of the bubble radius $R$, alongside the definition of the unstable equilibrium radius $R_{\text{Ue}}(t)$ for different pressure conditions. The blue line corresponds to the \textit{stable} equilibrium radii and the red line corresponds to the \textit{unstable} equilibrium radii of the bubble. {The minimum of the liquid pressure $p_\text{C,e}(t)$ and the corresponding radius $R_\text{C,e}(t)$ are time-dependent, due to the finite wall velocity $\dot{R}(t)$ in Eq.~\eqref{eq:pL}. If the evolution of the bubble radius is a quasi-static process with $\dot R \rightarrow 0$, $R_{\text{C,e}}(t) \rightarrow R_\text{C}$ as given by Eq.~\eqref{eq:R_C} and $p_{\text{C,e}}(t) \rightarrow p_\text{C}$ as given by Eq.~\eqref{eq:p_C}.}}
    \label{fig:R_Ue}
\end{figure}

{While the Blake threshold presented in the previous section can predict the critical tension required for the onset of cavitation under quasi-static conditions, it does not account for dynamic effects. Especially, the viscous term $-4\mu\dot{R}(t)/R(t)$ in Eq.~\eqref{eq:pL} induces time-dependent changes in the shape of $p_\text{L}$ due to the variations of the wall velocity $\dot{R}(t)$, resulting in a transient Blake radius, $R_\text{C,e}(t)$, and a transient Blake threshold, $p_\text{C,e}(t)$.}
Under transient conditions, the bubble expands rapidly and, {as a result of its finite inertia,} its radius may thereby cross the stable area in Figure \ref{fig:R_Ue} to arrive at the red line, known as the \textit{unstable equilibrium radius}. From a thermodynamic viewpoint, the unstable equilibrium radius is energetically less favourable than its stable (i.e.~smaller) counterpart for a given pressure. The unstable equilibrium radius converges to the Blake radius for a sufficiently slow (i.e.~quasi-static) bubble expansion {and, hence, presents a clear boundary for the dynamic stability of an expanding bubble that is compatible with the established quasi-static theory.} We hence define, as previously considered by \citet{Ida2009} for multi-bubble systems, the unstable equilibrium radius $R_{\text{Ue}}(t)$ as the threshold for cavitation. 

Depending on the value of $p_{\infty}(t)$, three scenarios ought to be considered, as illustrated in Figure \ref{fig:R_Ue}. Cavitation does not occur and $R_{\text{Ue}}(t)\rightarrow \infty$ for $p_{\infty}(t) \ge 0$, while $R_{\text{Ue}}(t) = R_{\text{C,e}}(t)$ if the ambient pressure is $p_{\infty}(t) \le p_{\text{C,e}}(t)$. Note that $p_{\text{C,e}}(t)$ is the minimum stable liquid pressure computed with the instantaneous bubble wall velocity $\dot{R}(t)$, where $R_{\text{C,e}}(t)$ is the corresponding radius, contrary to the standard Blake threshold that assumes $\dot{R} = 0$. If $p_{\text{C,e}}(t) < p_{\infty}(t) < 0$, the unstable equilibrium radius $R_{\text{Ue}}(t)$ follows from Eq.~\eqref{eq:pL} and is obtained by solving
\begin{equation}
        p_{\text{G},0} \left( \frac{R_0}{R^\star(t)} \right)^{3\kappa} - \frac{2\sigma}{R^\star(t)} - 4\mu \frac{\dot{R}(t)}{R^\star(t)} - p_{\infty}(t) = 0,
        \label{eq:R_Ue_equation}
\end{equation}
{a polynomial with two real roots $R^\star$ that correspond to the stable equilibrium radius $R_\text{Se}$ and the unstable equilibrium radius $R_\text{Ue}$. These equilibrium radii can be easily distinguished with the help of the Blake radius, as the stable equilibrium radius is smaller than the Blake radius, whereas the unstable equilibrium radius is larger than the Blake radius. Eq.~\eqref{eq:R_Ue_equation} is a time-dependent polynomial with potentially non-integer exponents (dependent on $\kappa$) that renders an analytical solution intractable. Nevertheless, the unstable equilibrium radius $R_\text{Ue}(t)  = R^\star \geq R_\text{C,e}(t)$ can be easily determined from Eq.~\eqref{eq:R_Ue_equation} using conventional numerical root-finding methods, such as the simple bisection method applied in this study.}

A bubble is in an unstable state as long as $R(t) > R_{\text{Ue}}(t)$, signifying cavitation. 
Contrary to the Blake threshold, this cavitation threshold based on the unstable equilibrium radius does not assume a quasi-static expansion of the bubble, {accounts for the influence of a transient ambient pressure ($p_{\infty}(t)$) as well as the influence of a finite expansion rate ($\dot{R}(t) \neq 0$) of the bubble,} and may include other phenomena by extending the definition of Eq.~\eqref{eq:pL}.

\subsection{{Computational workflow}}

\begin{figure}
    \centering
    \includegraphics[width=0.65\linewidth]{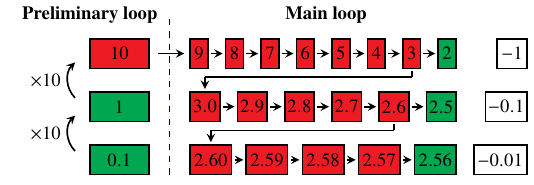}
    \caption{{Illustration of the procedure used to determine the critical tension ratio for cavitation onset, $p_\text{ng,C}/p_\text{C}$, for an arbitrary case where $p_\text{ng,C}/p_\text{C} = 2.56$. Each green and red rectangle corresponds to a computation where the applied tension ratio $p_\text{ng}/p_\text{C}$ is explicitly given inside each rectangle. Red rectangles refer to computations where cavitation occurs, i.e.~at some point during the expansion of the bubble $R(t) > R_\text{Ue}(t)$, while green rectangles correspond to computations where cavitation onset is not observed. The white rectangles correspond to the value of the increment used between each consecutive computation of $p_\text{ng,C}$.}}
    \label{fig:criticaltension_loop}
\end{figure}

{The considered configuration of a single gas bubble in a liquid excited by the single tension pulse presented in Section \ref{sec:methodology} is fully characterized by 9 parameters: $R_0$, $p_0$, $\rho$, $\sigma$, $\mu$, $c$, $\kappa$, $\tau$ and $p_\text{ng}$. To determine if the applied tension $p_\text{ng}$ is sufficient for cavitation onset,
the instantaneous bubble radius $R(t_{n})$ and bubble wall velocity $\dot{R}(t_{n})$ are obtained from 
the solution of Eq.~\eqref{eq:KM} at each discrete time instance $t_{n}$.
With $\dot{R}(t_{n})$, the instantaneous value of the unstable equilibrium radius $R_\text{Ue}(t_{n})$ is computed from Eq.~\eqref{eq:R_Ue_equation} using a bisection method.
The condition for cavitation is met when $R(t_{n}) > R_\text{Ue}(t_{n})$. 

In order to determine the critical tension $p_\text{ng,C}$ required for cavitation onset, we follow the procedure illustrated in Figure \ref{fig:criticaltension_loop}. We first conduct a preliminary loop whereby the amplitude of the tension pulse is increased by factors of 10 relative to the Blake pressure $p_\text{C}$, until cavitation is observed. Subsequently, the main loop works its way through successively decreasing tension ratios, until the critical tension is defined with 1 \% tolerance relative to the Blake pressure $p_\text{C}$.} {Consequently, for a single bubble system defined by the 9 parameters mentioned above, the critical tension for cavitation onset, $p_\text{ng,C}$, is defined as the largest tension (or, equivalently, the smallest pressure) at which cavitation is \textit{not} observed, meaning that a tension above this value is sufficient for the bubble radius to satisfy $R(t) > R_\text{Ue}(t)$ at some point during its expansion.}

\subsection{{Validation}}

\begin{figure}[t]
    \centering
    \includegraphics[width=\linewidth]{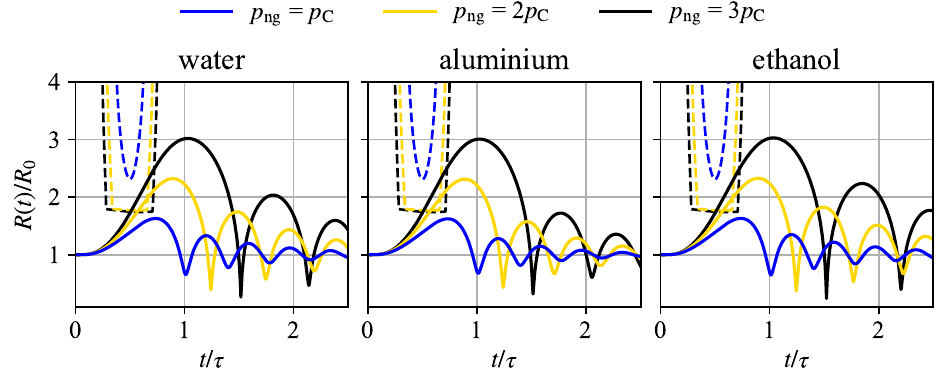}
    \caption{Evolution of the normalized radius $R(t)/R_{0}$ (solid lines) and the normalized unstable equilibrium radius $R_{\text{Ue}}(t)/R_{0}$ (dashed lines) for a bubble subjected to a single tension pulse, Eq.~\eqref{eq:pulse}, of duration $\tau = 10 t_{\text{i}}$.}
    \label{fig:onset_preliminary_pulse}
\end{figure}

{In order to validate the cavitation threshold proposed in Section \ref{sec:newthreshold}, we consider a single bubble in water, aluminium and ethanol that is subject to the tension pulse defined in Eq.~\eqref{eq:pulse}, with duration $\tau = 10t_{\text{i}}$, shown in Figure \ref{fig:onset_preliminary_pulse}. Note that, since the liquid is only temporarily under tension, cavitation onset is also transient.
The bubble radius (solid lines) grows beyond the unstable equilibrium radius $R_\text{Ue}(t)$ (dashed lines), signifying cavitation, for tension ratios in the range $1 < p_{\text{ng}}/p_{\text{C}}< 2$ in all three liquids. In this example, we choose the initial size and ambient pressure of the bubbles in the three different liquids such that they grow to the same maximum radius for $p_\text{ng} = p_\text{C}$. As a consequence, cavitation occurs at similar tension ratios in all three liquids and the bubble radius evolution remains qualitatively identical, suggesting that a self-similar description of cavitation onset is possible.}

\begin{figure}[t]
    \centering
    \includegraphics[width=\linewidth]{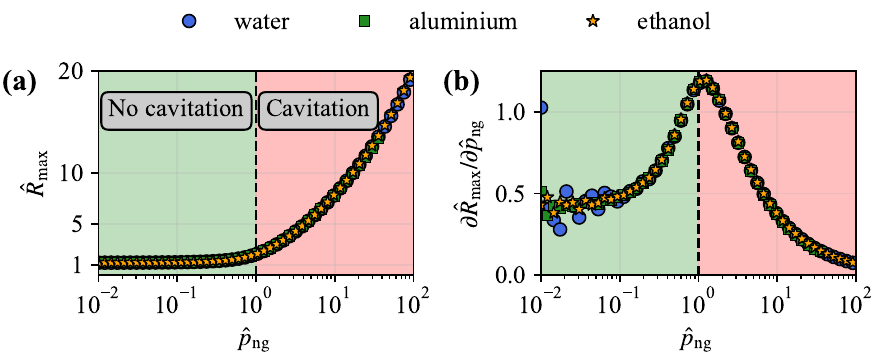}
    \caption{(a) Maximum normalized radius $\hat{R}_\text{max} = R_\text{max}/R_{0}$ as a function of the applied normalized liquid tension $\hat{p}_{\text{ng}} = p_{\text{ng}}/p_{\text{ng,C}}$, where $p_{\text{ng,C}}$ stands for the critical tension for cavitation onset, for a bubble subjected to a single tension pulse, Eq.~\eqref{eq:pulse}, of duration $\tau = 10 t_{\text{i}}$. (b) Partial derivative of $\hat{R}_\text{max}$ with regards to $\hat{p}_{\text{ng}}$ as a function of $\hat{p}_{\text{ng}}$.}
    \label{fig:onset_preliminary_pulse_Rmax}
\end{figure}

Identifying $p_{\text{ng,C}}$ as the critical tension for cavitation onset, i.e.~the bubble radius $R(t)$ surpasses $R_\text{Ue}(t)$ during the expansion of the bubble if $\hat{p}_{\text{ng}} = p_{\text{ng}}/p_{\text{ng,C}} > 1$, Figure \ref{fig:onset_preliminary_pulse_Rmax} displays how the normalized maximum radius $\hat{R}_{\text{max}} = R_\text{max}/R_{0}$ reached by the bubble correlates with the normalized liquid tension $\hat{p}_{\text{ng}}$. We observe that $\hat{R}_\text{max}$ increases noticeably when $\hat{p}_{\text{ng}} > 1$, see Figure \ref{fig:onset_preliminary_pulse_Rmax}(a), and that the proposed threshold for the onset of cavitation is closely aligned with the maximum increase in bubble radius, see Figure \ref{fig:onset_preliminary_pulse_Rmax}(b). 
{Hence, the proposed cavitation threshold coincides with a maximum rate of change of the maximum bubble radius with respect to the applied tension, demonstrating that the proposed threshold for cavitation corresponds to a characteristic feature of the bubble response.}
Furthermore, the evolution of $\hat{R}_\text{max}$ is the same for the three considered liquids. Note that the noise seen in Figure \ref{fig:onset_preliminary_pulse_Rmax}(b) for $\hat{p}_{\text{ng}} < 0.1$ is the result of numerical aliasing when computing the derivative of the data shown in Figure \ref{fig:onset_preliminary_pulse_Rmax}(a). 

In summary, these results support the criterion $R(t) > R_{\text{Ue}}(t)$ as a valid indicator of cavitation onset and suggest an underlying self-similarity. 
{The critical tension $p_{\text{ng,C}}$ required for the onset cavitation is defined as the tension for which the bubble grows to a radius of $R(t)=R_\text{Ue}(t)$, such that $p_{\text{ng}}/p_{\text{ng,C}}>1$ induces cavitation by pushing the bubble radius into the unstable regime where $R(t) > R_{\text{Ue}}(t)$.}

\section{Cavitation onset in different liquids}
\label{sec:onset}

\begin{figure}
    \centering
    \includegraphics[width=\linewidth]{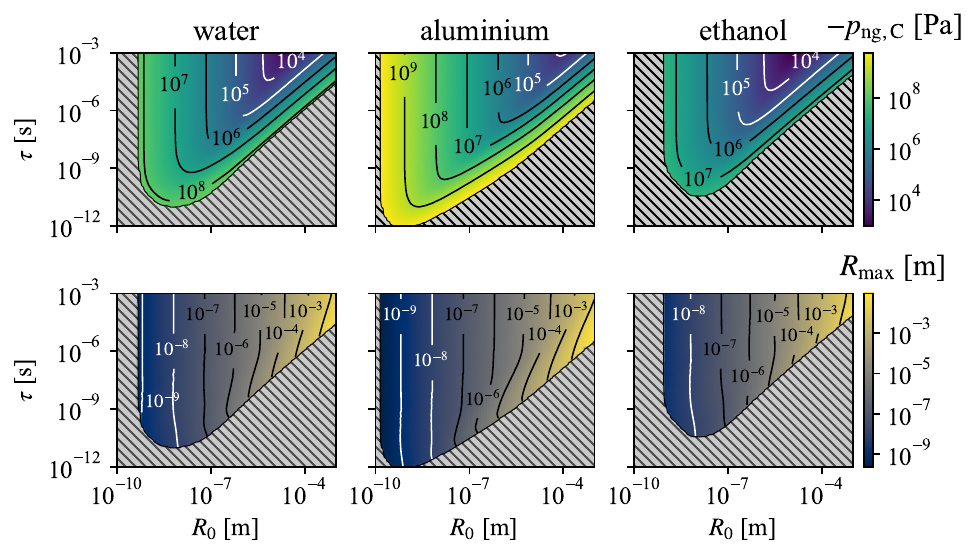}
    \caption{The critical tension for cavitation onset $-p_{\text{ng,C}}$ (top) and the maximum radius $R_{\text{max}}$ reached when applying $p_{\text{ng}}=p_{\text{ng,C}}$ (bottom), for an air bubble subjected to a single tension pulse, Eq.~\eqref{eq:pulse}, depending on $R_{0}$ and $\tau$ in different liquids. The hatched areas refer to $p_{\text{ng,C}} < p_{\text{frac}}$, Eq.~\eqref{eq:p_frac}. {The vertical isobars in the top figures correspond to $p_{\text{ng,C}} = p_{\text{C}}$, i.e.~the Blake threshold predicts the onset of cavitation, see Section \ref{sec:sub_Blake}.}}
    \label{fig:absolute_values}
\end{figure}

{We apply the methodology described in Section \ref{sec:methodology} to determine the critical tension $p_\text{ng,C}$ required for the onset of cavitation as proposed in Section \ref{sec:newthreshold} for air bubbles ($\kappa = 1.4$) with initial radii $R_0 \in [10^{-10} \, \text{m}, 10^{-3}\, \text{m}] $ that are subjected to the tension pulse defined in Eq.~\eqref{eq:pulse} with duration $\tau \in [10^{-12} \, \text{s} , 10^{-3}\, \text{s}]$ in water, aluminium, and ethanol (see Section \ref{sec:methodology} for the liquid properties). As the physical limit for the tension that can be sustained by a pure liquid, we consider the fracture pressure $p_\text{frac}$} estimated by nucleation theory at temperature $T = 300$ K and fracture time $t_{\text{frac}} = 10^{-6}$ s as \citep{Fisher1948}
\begin{equation}
    p_\text{frac} = -\sqrt{\frac{16\pi}{3kT} \frac{\sigma^3}{\ln (N t_{\text{frac}} k T /h )}},
    \label{eq:p_frac}
\end{equation}
where $k$, $h$ and $N$ are the Boltzmann, Planck and Avogadro constants, respectively.
Figure \ref{fig:absolute_values} displays the evolution of the critical tension $-p_{\text{ng,C}}$ and the corresponding maximum radius $R_{\text{max}}$ of an air bubble, depending on the initial radius $R_{0}$ and the duration of the tension pulse $\tau$ in the three considered liquids. The hatched areas in Figure \ref{fig:absolute_values} refer to $p_{\text{ng,C}} < p_{\text{frac}}$, where $p_{\text{frac}}$ is the fracture pressure of the liquid. 

\begin{figure}
    \centering
    \includegraphics[width=\linewidth]{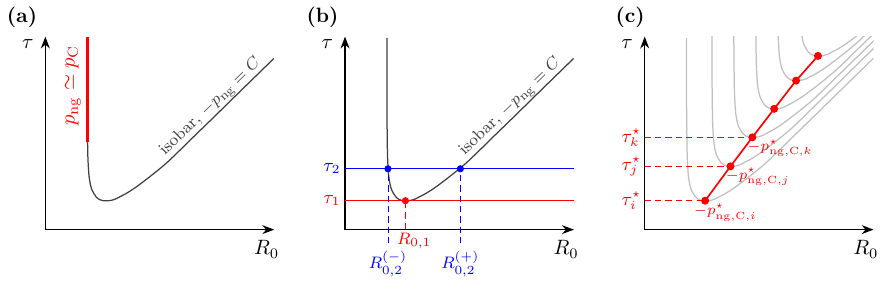}
    \caption{{Illustration of the information obtained from the phase maps shown in the top row of Figure \ref{fig:absolute_values}. (a) Vertical isobars correspond to $p_{\text{ng,C}} = p_{\text{C}}$, indicating that the Blake threshold predicts the onset of cavitation, see Section \ref{sec:sub_Blake}. (b) For a tension pulse with amplitude $-p_\text{ng,C}=C$, where $C$ is a positive constant, bubbles of a single initial radius $R_{0,1}$ cavitate if the tension pulse has duration $\tau_1$, whereas bubbles with an initial size in the range $R_0 \in [R_{0,2}^{(-)},R_{0,2}^{(+)}]$ cavitate if the tension pulse has duration $\tau_2>\tau_1$. (c) Minimum critical tension $-p_\text{ng,C}^\star$ required for the onset of cavitation for a corresponding pulse duration $\tau^\star$, where the red line represents the locus of this minimal tension, which is analyzed in Figure \ref{fig:p-tau}.}}
    \label{fig:plot_explanation}
\end{figure}

We can readily observe in Figure \ref{fig:absolute_values} that for a given tension, only bubbles in a certain size range can cavitate, and only if the tension pulse is of sufficiently long duration. This allows a direct assessment of cavitation under specific pressure conditions induced by a flow or by an external pressure source. 
{For a given bubble size, the minimum tension required for cavitation onset is the Blake threshold, see Eq.~\eqref{eq:p_C}, which is signified in the $R_0$-$\tau$ graph of Figure \ref{fig:absolute_values} by vertical isobars, as highlighted in Figure \ref{fig:plot_explanation}(a).}
The maximum radius achieved at cavitation onset is $R_\text{max} = O(R_0)$ and nearly independent of $\tau$, explaining the frequently postulated thresholds of $2 R_0 \lesssim R_\text{crit} \lesssim 3.5 R_0$ for cavitation onset \citep{Flynn1975a,Chomas2001}.

\subsection{{Influence of the pulse duration}}

\begin{figure}[t]
    \centering
    \includegraphics[width=\linewidth]{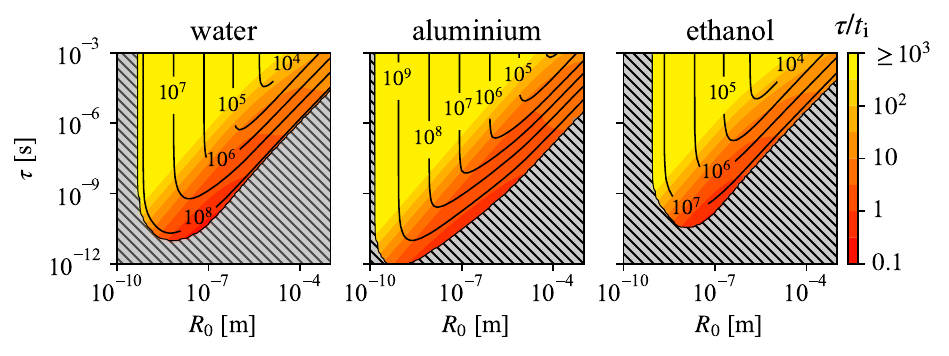}
    \caption{The timescale ratio $\tau/t_{\text{i}}$ for an air bubble subjected to a single tension pulse, Eq.~\eqref{eq:pulse}, depending on $R_{0}$ and $\tau$ in different liquids. The corresponding contour values of $-p_{\text{ng,C}}$ (Pa) are displayed on top, see the top row in Figure \ref{fig:absolute_values}. The hatched areas refer to $p_{\text{ng,C}} < p_{\text{frac}}$, Eq.~\eqref{eq:p_frac}.}
    \label{fig:absolute_values_tauti}
\end{figure}

Figure \ref{fig:absolute_values_tauti} displays the timescale ratio $\tau/t_{\text{i}}$ corresponding to a given pair of $R_{0}$ and $\tau$, with contours of $-p_{\text{ng,C}}$ plotted on top. The critical tension $-p_{\text{ng,C}}$ increases as $R_{0}$ increases for $\tau/t_{\text{i}} \lesssim 10$, since larger bubbles have a larger inertial timescale $t_{\text{i}}$, see Eq.~\eqref{eq:t_i}, meaning they respond slower to a change in pressure compared to their smaller counterparts; a larger tension is required for cavitation onset with a pulse of shorter duration $\tau/t_{\text{i}}$.  Moreover, $-p_{\text{ng,C}}$ decreases as $\tau$ increases. As long as $\tau/t_{\text{i}} \gtrsim 10^{3}$, $p_{\text{ng,C}}$ is virtually unaffected by $\tau$ and equal to the Blake threshold $p_{\text{C}}$, see Eq.~\eqref{eq:p_C}, which is signified in the $R_0$-$\tau$ graph of Figure \ref{fig:absolute_values} by vertical isobars,  {see Figure \ref{fig:plot_explanation}(a).
Hence, for a sufficiently long tension pulse, the bubble expansion is sufficiently slow for the quasi-static assumption to hold in good approximation and the Blake threshold marks the minimum tension required for the onset of cavitation.} 

These results explain previous observations regarding the accuracy of the Blake threshold reported in  experimental and numerical studies on nano- and microbubbles driven by ultrasound at subresonance frequencies \citep{Ilovitsh2018, Guemmer2021, Mancia2021}, since a bubble with $R_0 = O(10^{-9})$~m or $R_0 = O(10^{-6})$~m excited at subresonance frequencies of $f_\text{a} = O(10^{6})$~Hz or $f_\text{a} = O(10^{5})$~Hz, respectively, is in the regime where the Blake threshold $p_{\text{C}}$ marks the onset of cavitation.

\subsection{Relationship between minimum tension and pulse duration}
\label{sec:mintension}

{Each isobar shown in Figure \ref{fig:absolute_values} is formally defined as $-p_\text{ng}(R_0,\tau) = C$ for some positive constant $C$ and is observed to have a minimum where $\nabla p_\text{ng}(R_0,\tau) = 0$ and the Hessian of $-p_\text{ng}(R_0,\tau)$ is positive definite. This indicates that for any pulse duration $\tau^\star$, there exists a minimum critical tension $-p_\text{ng,C}^\star$ that is required for cavitation.
Applying a pulse of duration $\tau_1=\tau^\star$ and the corresponding pressure amplitude $p_\text{ng,C}^\star$, only bubbles of a single initial radius $R_{0,1}$ cavitate, as illustrated in Figure \ref{fig:plot_explanation}(b)}.
If the applied tension pulse has the same duration but with an amplitude larger than this minimum tension {or, similarly, the tension pulse increases in duration but maintains the same amplitude, see Figure \ref{fig:plot_explanation}(b)}, bubbles in a specific size range may cavitate, {which corresponds to the initial radii enveloped by the isobar $-p_\text{ng}(R_0,\tau) = C$.}

\begin{figure}[t]
    \centering
    \includegraphics[width=\linewidth]{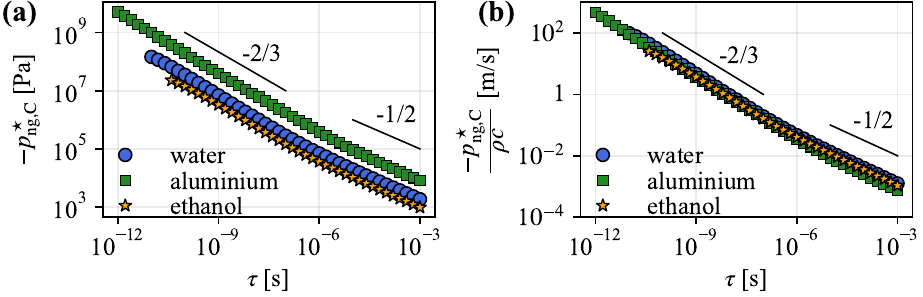}
    \caption{(a) Minimum critical tension $-p_\text{ng,C}^\star$ required for the onset of cavitation of an air bubble subjected to a single tension pulse, Eq.~\eqref{eq:pulse}, of a given duration $\tau$ in different liquids. (b) Minimum critical tension $-p_\text{ng,C}^\star$ divided by the specific acoustic impedance $\rho c$.}
    \label{fig:p-tau}
\end{figure}

{Figure \ref{fig:p-tau}(a) displays the locus of  the minimum tension $-p_\text{ng,C}^\star$ required for the onset of cavitation as a function of the pulse duration $\tau$, see Figure \ref{fig:plot_explanation}(c).}
We observe a power-law relationship {$-p_\text{ng,C}^\star \propto \tau^{-\beta}$, with $1/2 \lesssim \beta \lesssim 2/3$, or equivalently} $-p_\text{ng,C}^\star \propto \tau^{-1/\alpha}$, with $1.5 \lesssim \alpha \lesssim 2$. This is consistent with previous studies in the context of medical ultrasound, where the peak negative ultrasound pressure $p_\text{pnp}$ required for the onset of transient cavitation is related to the ultrasound frequency $f_\text{a}$ by $p_\text{pnp}^\alpha \propto f_\text{a}$, with $\alpha \in [1.67 , 2.1]$ \citep{Apfel1991} or $\alpha \in [1.16 , 1.56]$ \citep{Bader2013}.

Dividing $-p_\text{ng,C}^\star$ by the specific acoustic impedance $z = \rho c$, see Figure \ref{fig:p-tau}(b), the data of the three considered liquids collapse. The quantity $p/(\rho c)$ is proportional to the variation of the Helmholtz free energy $\Delta F \propto \Delta p^2/(\rho^2 c^2)$ of a bubble-containing liquid in a closed system \citep{Fuster2014a} and represents the particle velocity in classical acoustics. This indicates that the tension required for cavitation onset is directly related to the storage of elastic energy in the liquid \citep{Landau1959} and that, in order to induce cavitation, a shorter tension pulse requires a larger acoustically induced motion. {From a practical viewpoint, given that the data for $-p_\text{ng,C}^\star/(\rho c)$ with respect to the pulse duration $\tau$ coincide for different liquids, it is possible to relate data for the cavitation onset in one liquid to the cavitation onset in another liquid, simply based on the density and speed of sound of the liquids. This is valuable, for instance, for ultrasound applications and, more generally, determining cavitation in liquid metals, for which no optical access is easily possible in experiments \citep{Krivokorytov2018, Abramov2021}.}

\section{Self-similarity of cavitation onset}
\label{sec:results}

{The results shown in Figures \ref{fig:onset_preliminary_pulse}, \ref{fig:onset_preliminary_pulse_Rmax}, and \ref{fig:p-tau} indicate a self-similar response of a bubble to a transient tension event. In the following, we first define a dimensionless framework for the considered single-bubble system, which we then use to study the onset of caviation in dimensionless parameter space.}

\subsection{{Dimensionless framework}}
\label{sec:sub_framework}

The considered single-bubble system consists of 9 physical variables ($R_0$, $p_0$, $\rho$, $\sigma$, $\mu$, $c$, $\kappa$, $\tau$ and $p_\text{ng}$) and 3 dimensions (length, mass and time). Invoking the Buckingham $\pi$ theorem, this system can be described by $9-3=6$ dimensionless numbers. 
We describe the relevant physical phenomena governing the bubble behaviour (inertia, viscosity, and surface tension) by their respective characteristic timescales. {The inertia of pressure-driven bubble dynamics is readily characterized by the inertial timescale in Eq.~\eqref{eq:t_i}.}
To quantify the influence of viscosity, the viscous timescale is defined as $t_{\mu} = {\rho R_{0}^2}/{\mu}$,
which is representative of the time required for momentum to diffuse over a distance $R_{0}$ {and, in the context of pressure-driven bubble dynamics, represents the timescale over which the bubble wall motion is decelerated by viscous effects \citep{Franc2005}.} The effect of surface tension is taken into account through the capillary timescale, $t_{\sigma} = \sqrt{{\rho R_{0}^3}/{\sigma}}$,
which, {in general, is representative of the dispersion due to surface tension over a distance $R_{0}$ \citep{Denner2016} and, for bubble dynamics, is a measure of the collapse time of a bubble that is driven only by surface tension \citep{Franc2005}.
Using these timescales} alongside the Blake threshold $p_{\text{C}}$, the proposed dimensionless numbers are $\kappa$, $t_{\mu}/t_{\text{i}}$, $t_{\sigma}/t_{\text{i}}$, $\tau/t_{\text{i}}$, $p_{\text{ng}}/p_{\text{C}}$ and $-p_{\text{ng}}/K$, where $K = \rho c^2$ stands for the bulk modulus of the liquid. Preliminary studies have shown that the speed of sound does not have a marked influence on the self-similarity of the results, since the order of magnitude of $-p_{\text{ng}}/K$ is the same for typically encountered liquids.

The aim of this dimensionless framework is to obtain the critical tension ratio $p_{\text{ng,C}}/p_{\text{C}}$ above which cavitation is observed for a given set of dimensionless numbers ($t_{\mu}/t_{\text{i}}$, $t_{\sigma}/t_{\text{i}}$, $\tau/t_{\text{i}}$, $\kappa$), independent of the specific liquid or initial bubble size. 
The values for the dimensionless numbers considered in the following are $t_{\mu}/t_{\text{i}} \in [10^{-1}, 10^{5}]$, $t_{\sigma}/t_{\text{i}} \in [1.5, 10^{2}]$, $\tau/t_{\text{i}} \in \{10^{2}, 10^{3}, 10^{4}\}$, and $\kappa \in \{1.4, 1.67\}$, representing an air or helium bubble in water with an initial size of $R_{0} \in [10^{-9}\, \text{m},10^{-3}\, \text{m}]$, assuming a reference ambient pressure of $p_{0} = 10^{5}$ Pa.

\subsection{{Cavitation onset in dimensionless parameter space}}

\begin{figure}[t]
    \centering
    \includegraphics[width=\linewidth]{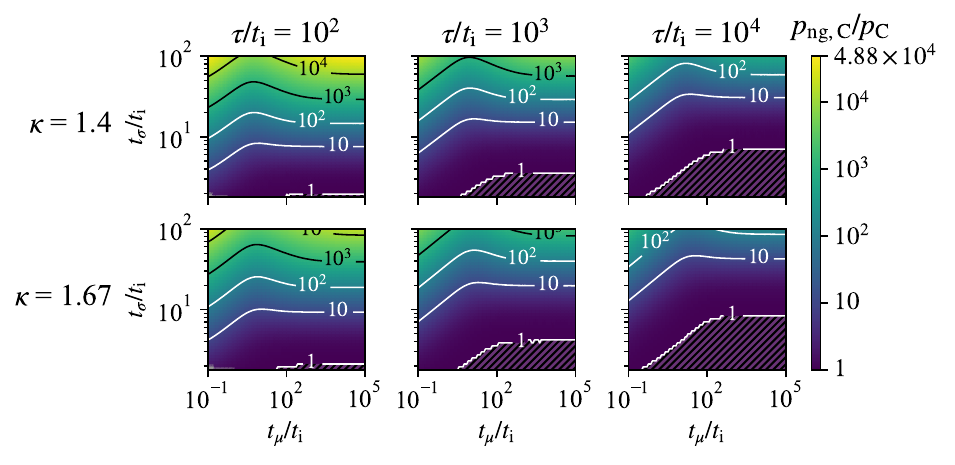}
    \caption{The critical tension ratio $p_{\text{ng,C}}/p_{\text{C}}$ required for the onset of cavitation of a bubble subjected to a single tension pulse, Eq.~\eqref{eq:pulse}, depending on $t_{\mu}/t_{\text{i}}$ and $t_{\sigma}/t_{\text{i}}$. The rows and columns correspond to certain values of $\kappa$ and $\tau/t_{\text{i}}$, respectively. Hatched areas refer to $p_{\text{ng,C}}/p_{\text{C}} = 1${, i.e.~the Blake threshold predicts the onset of cavitation, see Section \ref{sec:sub_Blake}}.}
    \label{fig:main_result}
\end{figure}

\begin{figure}[t]
    \centering
    \includegraphics[width=\linewidth]{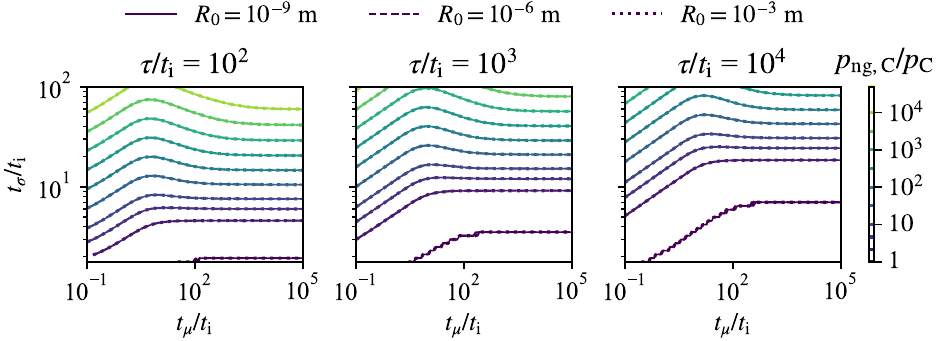}
    \caption{The same figure as Figure \ref{fig:main_result}, but displaying only the contour values of the critical tension ratio $p_{\text{ng,C}}/p_{\text{C}}$ for cavitation onset, for $\kappa = 1.4$. Each line style corresponds to a different {imposed} initial radius $R_{0}$. {For each value of $R_{0}$, the other physical parameters characterizing the single-bubble system ($p_{0}$, $\rho$, $\mu$, $\sigma$, $\tau$) are adjusted so as to have the same values for the dimensionless numbers ($t_{\mu}/t_{\text{i}}$, $t_{\sigma}/t_{\text{i}}$, $\tau/t_{\text{i}}$) and to validate the self-similarity description of cavitation onset.}}
    \label{fig:self_similarity}
\end{figure}

{Employing the dimensionless framework presented in the previous section,}
Figure \ref{fig:main_result} displays the critical tension ratio $p_{\text{ng},\text{C}}/p_{\text{C}}$ for cavitation onset with a single tension pulse, in function of the dimensionless numbers $\kappa$, $\tau/t_{\text{i}}$, $t_{\mu}/t_{\text{i}}$ and $t_{\sigma}/t_{\text{i}}$.
{The hatched areas correspond to $p_{\text{ng},\text{C}}/p_{\text{C}} = 1$, representing the set of dimensionless parameters ($\kappa$, $t_{\mu}/t_{\text{i}}$, $t_{\sigma}/t_{\text{i}}$, $\tau/t_{\text{i}}$) for which the standard quasi-static theory and the Blake threshold $p_\text{C}$, see Eq.~\eqref{eq:p_C}, predict the onset of cavitation, see Section \ref{sec:sub_Blake}. Additionally, we note that $p_{\text{ng},\text{C}}/p_{\text{C}} \ge 1$ throughout the considered parameter space, indicating that the Blake threshold $p_{\text{C}}$ is the minimum tension required for cavitation onset, as already observed in Section \ref{sec:onset}.} 
As $\tau/t_{\text{i}}$ increases, $p_{\text{ng},\text{C}}/p_{\text{C}}$ decreases and the hatched areas occupy a larger area, illustrating that the evolution of the bubble radius becomes closer to a quasi-static process. 
Therefore, the Blake threshold $p_{\text{C}}$ becomes the effective critical tension for a larger range of values ($\kappa$, $t_{\mu}/t_{\text{i}}$, $t_{\sigma}/t_{\text{i}}$). 
{Moreover, we observe a strong influence of surface tension on the critical tension required for cavitation, commensurate with its role as the main physical mechanism that renders a bubble stable in a liquid under tension.}
The critical tension ratio $p_{\text{ng},\text{C}}/p_{\text{C}}$ also decreases as $\kappa$ increases, since larger values of $\kappa$ (under the employed isentropic assumption) mean that the pressure in the gas decreases faster as the bubble expands. Surpassing $R_{\text{Ue}}(t)$ during the time in which $p_{\infty}(t) < 0$ is, thus, easier.

It is important to note that the results shown in Figures \ref{fig:main_result} are self-similar. For instance, when imposing a specific value for the initial bubble radius $R_{0}$, and adapting the other physical variables ($p_{0}$, $\rho$, $\mu$, $\sigma$, $\tau$) in order to match the desired set of dimensionless numbers ($t_{\mu}/t_{\text{i}}$, $t_{\sigma}/t_{\text{i}}$, $\tau/t_{\text{i}}$) {independent of the imposed value of $R_0$}, the results exactly collapse, as seen in Figure \ref{fig:self_similarity}.

\subsection{{The influence of viscosity}}

{In the classic cavitation theory based on the Blake threshold, the bubble expansion is assumed to proceed in a quasi-static manner ($\dot{R} \rightarrow 0$) and the influence of viscosity is, consequently, neglected. This is not the case if the transient nature of a real pressure field and the finite expansion rate of a bubble are considered.
In the considered case of a single tension pulse,} because the ambient pressure $p_{\infty}(t)$ recovers its initial value $p_{0}$ after the passage of the tension pulse, there is a competition between the time the bubble takes to expand beyond the unstable equilibrium radius $R_{\text{Ue}}(t)$ and the time for which {a sufficient tension applies}.
In the low-viscosity regime ($t_{\mu}/t_{\text{i}} \gtrsim 500$), the critical tension ratio $p_{\text{ng},\text{C}}/p_{\text{C}}$ increases as $t_{\sigma}/t_{\text{i}}$ increases. In fact, as $t_{\sigma}/t_{\text{i}}$ increases, $|p_{\text{C}}|/p_{0}$ decreases by several orders of magnitude and, in turn, the time interval for which $p_{\infty}(t) < 0$ decreases. Thus, cavitation onset requires $p_{\text{ng}}/p_{\text{C}} \gg 1$. 
In the high-viscosity regime ($t_{\mu}/t_{\text{i}} \lesssim 10$), $p_{\text{ng},\text{C}}/p_{\text{C}}$ increases as $t_{\mu}/t_{\text{i}}$ decreases, since the influence of viscosity is strong enough to slow down the expansion of the bubble. Therefore, to give the bubble sufficient time to expand beyond $R_{\text{Ue}}$, a larger critical tension ratio $p_{\text{ng},\text{C}}/p_{\text{C}}$ is required. In the intermediate-viscosity regime ($10 \lesssim t_{\mu}/t_{\text{i}} \lesssim 500$), $p_{\text{ng},\text{C}}/p_{\text{C}}$ decreases as $t_{\mu}/t_{\text{i}}$ decreases, since viscosity is not strong enough to slow down the expansion of the bubble. However, $R_{\text{Ue}}(t)$ is smaller compared to the low-viscosity regime ($t_{\mu}/t_{\text{i}} \gtrsim  500$), due to the influence of viscosity in Eq.~\eqref{eq:R_Ue_equation}, and cavitation onset is observed at smaller critical tension ratios.

\begin{figure}
    \centering
    \includegraphics[width=\linewidth]{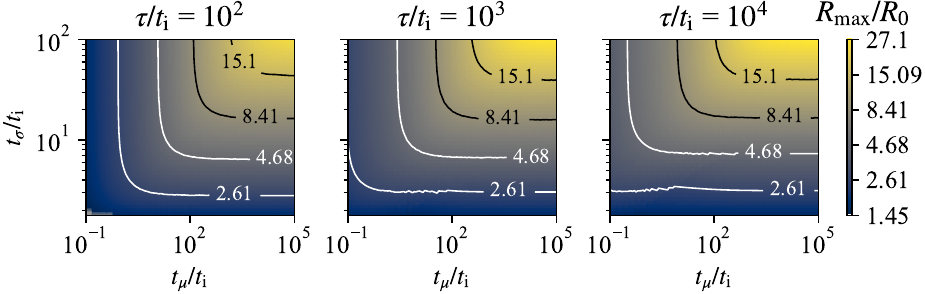}
    \caption{The normalized achieved maximum radius $R_{\text{max}}/R_{0}$ when applying $p_{\text{ng}} = p_{\text{ng,C}}$ for a bubble subjected to a single tension pulse, see Eq.~\eqref{eq:pulse}, for $\kappa = 1.4$.}
    \label{fig:max_Radius}
\end{figure}

The achieved maximum radius when applying $p_{\text{ng}} = p_{\text{ng,C}}$ is displayed, for $\kappa = 1.4$, in Figure \ref{fig:max_Radius}. We notice that $R_{\text{max}}/R_{0}$ increases with $t_{\sigma}/t_{\text{i}}$ until a certain threshold value of $t_{\mu}/t_{\text{i}}$. The threshold of $t_{\mu}/t_{\text{i}}$ highlights that as viscosity becomes dominant, the bubble expansion is slowed down and, therefore, the achieved maximum radius decreases.

\section{{Simplifying assumptions and limitations}}
\label{sec:simplifications}

{In order to define the proposed dynamic threshold for the onset of cavitation and study the conditions for cavitation in response to a transient pressure pulse, we have made three main simplifications: the bubble is spherical at all times, heat and mass transfer are neglected, and the tension pulse has a $\sin^2$-shape. Although such simplifications are common place for studying the fundamentals of pressure-driven bubble dynamics \citep{Brennen1995, Franc2005, Fuster2019}, they warrant a more detailed discussion.}

{The assumption of a spherical bubble neglects the variety of shapes a bubble can assume in a transient pressure field. However, the here considered spherical bubble can serve as a prototype for the behaviour of bubbles of any shape. This has been empirically demonstrated in previous experimental and computational studies \citep{Gonzalez-Avila2021,Peng2018}. Moreover Rayleigh-Plesset-type models, such as the Keller-Miksis model employed in this study, can also be derived for bubbles in cylindrical symmetry \citep{Denner2024a}, for which the liquid pressure at the gas liquid interface is given as  
\begin{equation}
    p_\mathrm{L} = p_\mathrm{G} - \frac{\sigma}{R} - 2 \mu \frac{\dot{R}}{R}, \label{eq:pL_cyl}
\end{equation}
from which the equation to determine the unstable equilibrium radius follows as
\begin{equation}
    p_{\text{G},0} \left( \frac{R_0}{R^\star(t)} \right)^{2\kappa} - \frac{\sigma}{R^\star(t)} - 2\mu \frac{\dot{R}(t)}{R^\star(t)} - p_{\infty}(t) = 0.
    \label{eq:R_Ue_cyl}
\end{equation}
Note the similarity between Eqs.~\eqref{eq:pL} and \eqref{eq:pL_cyl}, as well as Eqs.~\eqref{eq:R_Ue_equation} and \eqref{eq:R_Ue_cyl}, re-enforcing the notion that a spherical bubble is a suitable prototype for studying cavitation onset.}

{We further assumed that heat and mass transfer are negligible and, as a consequence, the vapour pressure is $p_\text{v}=0$. What may seem to be a severe simplification with respect to cavitation, a phenomenon that is often induced by phase change, must be seen in the context of the timescales considered in this study. The considered single tension pulse has a duration of $\tau \leq 10^{-3} \, \text{s}$, which is representative of the pressure timescales driving cavitation in medical ultrasound \citep{Wan2015}, sonochemistry applications \citep{Meroni2022}, and impulsively driven liquids \citep{Sobral2024}, but which is too short for heat and mass transfer to play an important role. Incorporating a finite vapour pressure would merely entail an addition to the right-hand side of Eq.~\eqref{eq:pL} and, accordingly, to Eq.~\eqref{eq:R_Ue_equation}, which can be readily accounted for in the presented approach to determine cavitation. The unstable equilibrium radius is $R_\text{Ue} \rightarrow \infty$ for $p_\text{L} \rightarrow p_\text{v}$, which consequently shifts the stable area marginally to larger pressures in Figure \ref{fig:p_L}. However, considering the small amplitude of the vapour pressure compared to the critical tension required for cavitation onset of most engineering liquids, such as water and ethanol with $p_\text{v} = \mathcal{O}(10^3) \, \text{Pa}$ and liquid metals with $p_\text{v} < 1 \, \text{Pa}$, assuming that $p_\text{v} = 0$ does not have a major impact on the presented results. 
We refer the reader to the textbook of \citet{Franc2005}, especially Section 2.2 therein, for an excellent discussion on the influence of a finite vapour pressure.}

{In this study, we model a transient pressure field by a single tension pulse with a $\sin^2$-shape. This study shows that for cavitation to occur, the essential condition is that the bubble receives enough acoustic energy to expand beyond its unstable equilibrium radius. 
In this sense, the detailed shape of the pressure transient is secondary: what matters is that the combination of amplitude and duration carries enough acoustic energy to drive a sufficient growth of the bubble. To this end, the area-specific acoustic energy of a planar pressure pulse of duration $\tau$ is given as
\begin{equation}
    e_{\tau} = \int_0^\tau \frac{p_\text{ng}^2}{\rho c} \, \text{dt}.
\end{equation} 
Using a smooth, compactly supported prototype pulse such as a $\sin^2$-profile captures this energy input while avoiding artificial features of more arbitrary waveforms. If the pressure transient has a slightly different shape, the primary adjustment needed is to rescale the pulse amplitude or duration so that the effective energy delivered to the bubble is the same. Thus, a single $\sin^2$-pulse can serve as a suitable surrogate for transient pressure fields for the purposes of studying cavitation onset.}

\reviewertwo{Lastly, we note that our preliminary results did not show a noticeable influence of the compressibility of the liquid on the proposed cavitation threshold, as mentioned in Section \ref{sec:sub_framework}, neither with the Rayleigh-Plesset equation that assumes an incompressible liquid \citep{Plesset1977}, nor with the Gilmore model that assumes a compressible liquid with pressure-dependent properties \citep{Gilmore1952}. However, accounting for the compressibility of the liquid allows to establish a relationship between the minimum tension required for cavitation onset and the storage of elastic energy in the liquid, discussed in Section \ref{sec:mintension}. Therefore, we chose the Keller-Miksis model in Eq.~\eqref{eq:KM} for this study, as it accounts for all physical mechanisms relevant to the onset of cavitation. 
% in order to be able to conduct a comprehensive analysis, such as the scaling shown in Figure \ref{fig:p-tau}(b).
}

\section{Conclusions}
\label{sec:conclusions}

Based on the hypothesis that cavitation is marked by the transition of the radius of a pressure-driven bubble to a dynamically unstable state, we have identified the unstable equilibrium radius as the dynamic threshold for cavitation in transient pressure fields. This threshold is based on first principles and depends on the instantaneous state of the gas-liquid system. With this new cavitation threshold, a methodology based upon a dimensionless framework has been proposed to highlight that cavitation onset is self-similar, considering a single tension pulse as a representative transient pressure field. Phase maps for different liquids reveal a minimum tension that is required for cavitation onset, which is determined by the duration of the tension event and the initial bubble size, and described by a power-law relationship. For sufficiently long tension events, the well-known Blake threshold emerges as the lower bound for cavitation across all considered conditions. Moreover, the bubble radius at cavitation onset is nearly independent of the duration of liquid tension, explaining the origin of previously proposed cavitation thresholds based on the bubble radius.

This work provides a unifying, physics-based criterion for cavitation under transient conditions, bridging the gap between quasi-static and dynamic regimes and offering a predictive framework for cavitation onset. To this end, the code developed for quantifying cavitation onset is openly available in the accompanying Zenodo repository (see the data availability statement). While the current study has been based on certain assumptions (e.g.~mass transfer is neglected), 
the proposed method for the identification and quantification of cavitation onset can be readily extended to include additional physical phenomena.

\section*{Acknowledgements}
\noindent We acknowledge the support of the Natural Sciences and Engineering Research Council of Canada (NSERC), funding reference number RGPIN-2024-04805. 

\section*{Data availability}
\noindent The data that support the findings of this study as well as code to reproduce the presented results are openly available in Zenodo at \url{https://doi.org/10.5281/zenodo.15839418}.

% \bibliographystyle{model2-names}
% \bibliography{/Users/fabian/Zotero/library}

\end{document}